\documentstyle[emulateapj]{article}
\begin{document}
%\special{landscape}
\lefthead{IRWIN \& BREGMAN}
\righthead{X-RAY BULGE OF M31}

\slugcomment{Astrophysical Journal, accepted}

\title{Using the Bulge of M31 as a Template for the Integrated X-ray Emission
from LMXBs}

\author{Jimmy A. Irwin\altaffilmark{1} and Joel N. Bregman}
\affil{Department of Astronomy, University of Michigan, \\
Ann Arbor, MI 48109-1090 \\
E-mail: jirwin@astro.lsa.umich.edu, jbregman@umich.edu}

\altaffiltext{1}{Chandra (AXAF) Fellow.}

\begin{abstract}
We have performed joint {\it ASCA} + {\it ROSAT} PSPC spectral fitting on
the inner 5$^{\prime}$ of the bulge of M31. We find that single component
spectral models provide an inadequate fit to the spectrum, in contrast
to previous studies by {\it Einstein} and {\it Ginga}. Although the 2--10
keV spectrum can be fit adequately with a bremsstrahlung model with
$kT=7.4 \pm 0.3$ keV, an additional soft component
with $kT= 0.38 \pm 0.03$ keV is required to fit the
spectrum below 2 keV. This soft component comprises 38\% $\pm$ 6\% of the
total emission in the 0.1--2.0 keV band, and possibly more depending on
the absorption value used in the fit. Since previous spatial studies of the
bulge of M31 indicate that $<$ 25\% of the X-ray emission
from the bulge in this band is from a diffuse gaseous component,
this implies that stellar sources, namely low-mass X-ray binaries (LMXBs),
are responsible for some if not all of the
soft component. The spectral properties of M31 are very similar to those
of the X-ray faint early-type galaxy NGC~4382. This supports the claim that
the unexplained soft X-ray emission seen previously in these galaxies also
emanates from LMXBs.
\end{abstract}


\section{Introduction} \label{sec:intro}

Nearly all that is known about the X-ray spectra of low mass X-ray binaries
(LMXBs) has come from observations of LMXBs in the Galactic bulge or those
that lie in the Galactic plane. Unfortunately, this also
means that large quantities of Galactic hydrogen gas
($N_H \sim 10^{22}$ cm$^{-2}$) lie between us and the LMXBs we wish to observe.
Nearly all of the X-ray flux below 1 keV from these LMXBs is absorbed
by intervening material between us and the LMXB.
As a consequence, very little is known about the
X-ray properties of LMXBs at very soft X-ray energies.
A recent survey of 49 Galactic LMXBs observed with the {\it Einstein}
Observatory found that a majority of the spectra were adequately fit with a
powerlaw plus high energy exponential cutoff spectral model, with
$\Gamma$ between --0.2 and 1.0 and a high energy cutoff in the 3--7 keV range.
(Christian \& Swank 1997).
Thermal bremsstrahlung models with
$kT=5-10$ keV are also
frequently employed to describe the emission from LMXBs.
Such models contribute relatively little to the X-ray emission in
the 0.1--1 keV range compared to the 1--10 keV range. Given the large
hydrogen column densities towards most these objects, though, any soft
component would have been completely absorbed.

\begin{table*}[htb]
\label{tab:obs_info}
\begin{center}
{\sc Table~1}
\vskip0.1truein
\begin{tabular}{lcccc}
\multicolumn{5}{c}{\sc Observational Data Set For M31} \cr
\hline \hline
&Observation&&Energy Band& Net X-ray\\
Instrument&
Number &
Exposure (s)&
Used (keV)&
counts \\
\hline
{\it ROSAT} PSPC& RP600068N00 & 30,005 & 0.2--2.4 & 29,720 \\
{\it ASCA} GIS2 & 63007000 & 88,109 & 0.8--10.0 & 24,900 \\
{\it ASCA} GIS3 & 63007000 & 89,004 & 0.8--10.0 & 29,718 \\
{\it ASCA} SIS0 & 63007000 & 50,531 & 0.8--10.0 & 23,623 \\
{\it ASCA} SIS1 & 63007000 & 51,802 & 0.8--10.0 & 17,141 \\
\hline
\end{tabular}
\end{center}
\end{table*}

Is there reason to believe that the X-ray spectrum of LMXBs is interesting
below 1 keV?
There are two examples of Galactic LMXBs that lie in directions of low
hydrogen column densities that were observed with {\it ROSAT} and/or
{\it ASCA}. Both LMXBs show evidence for very soft X-ray emission.
Choi et al.\ (1997) confirmed earlier reports of a 0.1 keV blackbody
component in addition to a harder power-law component
in the X-ray spectrum of Her X-1 with {\it ASCA}.
The very soft blackbody component has been interpreted as the thermal
re-emission by an opaque distribution of gas around the neutron star.
The Galactic LMXB MS~1603+2600 also exhibits very soft emission
(Hakala et al.\ 1998), although there is some question as to whether
this system is an LMXB or a cataclysmic variable (Ergma \& Vilhu 1993).
There are several LMXBs in globular clusters that lie in directions
of low column densities that
do not show strong excess soft X-ray emission. However, these LMXBs reside
in low-metallicity environments. A recent study of 12 LMXBs located in
globular clusters of M31 by Irwin \& Bregman (1999) found a correlation between
the X-ray spectral properties of the LMXB with the metallicity of the host
globular cluster. The one LMXB in a globular cluster with greater than solar
metallicity had a much softer X-ray spectrum than those LMXBs located in
metal-poor clusters.

The possible existence of a soft component of LMXBs is particularly important
in the case of early-type galaxies that are very underluminous in X-rays for a
given optical luminosity (low $L_X/L_B$). In these galaxies it is suspected
that the X-ray emission is
primarily stellar in nature. Whereas it is well-established that
the X-ray emission in X-ray bright elliptical galaxies is predominantly
from hot ($\sim$0.8 keV) gas, X-ray faint galaxies appear to be
lacking this component. Instead, their X-ray emission is characterized by
a two-temperature (5--10 keV + 0.3 keV) model
(Fabbiano, Kim, \& Trinchieri 1994; Pellegrini 1994; Kim et al.\ 1996).
The 5--10 keV component is generally regarded as the integrated emission from
LMXBs, and has been seen in nearly all early-type galaxies observed with
{\it ASCA} (see, e.g., Matsumoto et al.\ 1997). The origin of the soft
component remains a mystery. Although it has been suggested that the source
of the emission might be warm interstellar gas,
recent work has suggested that the source of the soft emission is the same
collection of LMXBs responsible for the 5--10 keV component
(Irwin \& Sarazin 1998a,b).

Given the paucity of Galactic LMXB candidates that are not heavily absorbed,
the bulge of M31 provides the nearest laboratory for studying a large number
of LMXBs in high metallicity environments at low X-ray energies. At a Galactic
hydrogen column density of $6.7 \times 10^{20}$ cm$^{-2}$, the 0.1--1 keV
spectra of LMXBs in the bulge of M31 are not completely absorbed (25\%
transmission at 0.35 keV), as is the case towards the Galactic bulge. Here, we
analyze the joint {\it ASCA} and {\it ROSAT} PSPC spectrum of the inner
5$^{\prime}$ of the bulge of M31. By combining both instruments we are able to
fit the spectrum of the bulge of M31 over a broad energy range (0.2--10 keV),
using the advantages of both instruments to complement one another. This
spectrum can be used as a template for what the spectrum of a collection of
LMXBs should look like in more distant early-type galaxies.
 
\section{Previous X-ray Observations of the Bulge of M31} \label{previous_obs}

The first spectral study of the bulge of M31 was performed with the
{\it Einstein} IPC (0.2--4.5 keV) and MPC (1.2--10 keV) instruments by
Fabbiano, Trinchieri, \& Van Speybroeck (1987). For the inner 5$^{\prime}$ of
the bulge, this study found that a thermal bremsstrahlung model with
$kT= 6-13$ keV and no intrinsic absorption above the Galactic value fit the
data well. Makishima et al.\ (1989) used {\it Ginga} (2--20 keV) data
to study all of M31 (disk + bulge), and found a best-fit temperature of
$7.2 \pm 0.4$ keV with significant excess absorption, although this model
yielded a rather large reduced $\chi^2$ value. A better fit was
obtained with a powerlaw with high energy cutoff model with $\Gamma=1.9\pm0.3$
and a cutoff energy of $6.8\pm0.5$ keV. However, the absorption was once
again an order of magnitude higher than the Galactic value.

Supper et al.\ (1997) analyzed a long {\it ROSAT} PSPC observation of the
bulge of M31 and found a best-fit bremsstrahlung temperature of $\sim$1 keV,
although the authors state that the temperature could not be well-determined.
We interpret this to mean that the reduced $\chi^2$ value of this model
was large, and our analysis of the same data (\S~\ref{sec:spectral}) confirms
this. Irwin \& Sarazin (1998b) analyzed a much shorter (2800 second) {\it ROSAT}
PSPC observation and found a best-fit bremsstrahlung temperature of
$0.78 \pm 0.07$ keV from a fit that
was marginally acceptable, although a significantly better fit was found with a
two component model: a Raymond-Smith (1977) thermal model with
$kT=0.36^{+0.09}_{-0.06}$ keV and a metallicity
$0.012^{+0.012}_{-0.005}$ solar, and a harder bremsstrahlung component
with $kT>6.4$ keV. Both {\it ROSAT} PSPC results indicated that below
2 keV, the spectrum of the bulge of M31 was not well-represented by a
hard 5--10 keV bremsstrahlung model, in contrast to the {\it Einstein}
and {\it Ginga} results.

\begin{figure*}[htb]
\vskip4.1truein
\includegraphics{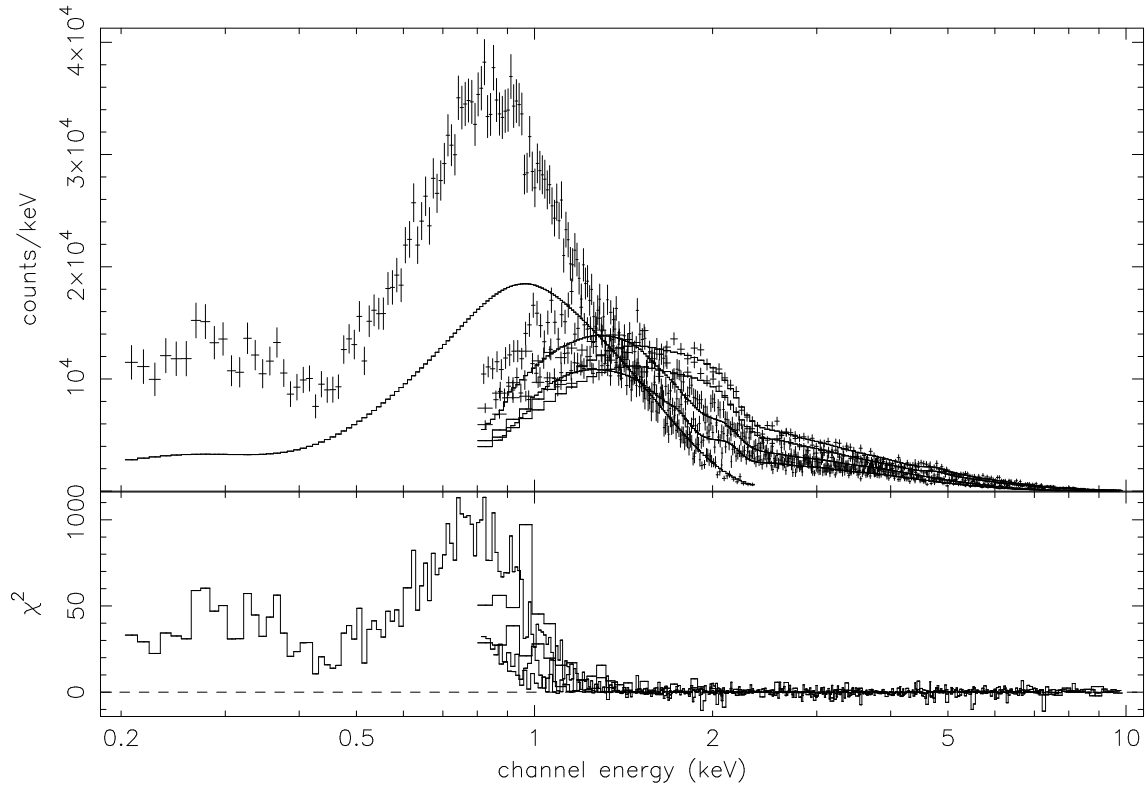}
\caption[TB only]{
{\it ROSAT} PSPC + {\it ASCA} data with best-fit one component TB
model fit to the 2.0--10 keV bandpass only. Data below 2.0 keV was then
added in after the fit. For all five instruments, an excess of soft emission
below 1.5 keV was detected.
\label{fig:brems}}
\end{figure*}

\section{{\it ROSAT} PSPC and {\it ASCA} Data Reduction} \label{sec:data}

We have chosen a long {\it ROSAT} PSPC observation of the bulge of M31 from
the HEASARC archive (RP600068N00). The exposure time was 30,005 seconds. The
spectrum of the inner 5$^{\prime}$ of the bulge was extracted, and the
energy channels were rebinned to contain at least 25 counts. A background
spectrum, extracted from an annulus of $30^{\prime}-40^{\prime}$ and corrected
for vignetting, was scaled to and subtracted from the source spectrum. Energy
channels below 0.2 keV and above 2.4 keV were then excluded.

A long {\it ASCA} observation of M31 was also taken from the HEASARC archive
(63007000). The data were screened using the standard screening criteria
applied to all the archival data (Revision 2 processing). The spectrum of
the inner 5$^{\prime}$ was extracted from the GIS2, GIS3, SIS0, and SIS1
data, with a total exposure time of 177,113 seconds for the combined GIS
data and 102,333 seconds for the combined SIS data. We chose to analyze the
BRIGHT2 SIS data, since the data can be corrected for echo and dark frame
error effects in this mode. The SIS data were taken
in 4-CCD mode, but nearly all of the inner 5$^{\prime}$ of the bulge fit
within one chip. Therefore, only data from this one chip were used in the
analysis, in order to avoid complications that might arise from averaging
together the responses from different chips.
Background was obtained from the deep blank sky data provided
by the {\it ASCA} Guest Observer Facility. We used the same region filter
to extract the background as we did the data, so that both background and
data were affected by the detector response in the same manner.
Energy channels below 0.8 keV and above 10 keV were excluded.
Once again, the energy channels were rebinned to contain at least 25 counts.
Because of differences in the calibrations among the five data sets, as well
as possible temporal variations in the flux between the {\it ROSAT} and
{\it ASCA} observations, we have chosen to let the normalizations of all
spectral models be free parameters. In all fits the SIS normalizations were
consistent with one another, but about 20\% less than the GIS and PSPC
normalization. This is a result of having excluded a fraction of the
data that fell off the primary chip.
Table~\ref{tab:obs_info} gives details of the observations.

\section{Results of Spectral Fitting} \label{sec:spectral}

\subsection{One Component Models} \label{ssec:one_comp}

As a first step, we have attempted to fit a variety of single component
spectral models to the data, the first of which being a thermal bremsstrahlung
(TB) model. A very poor fit to the data (reduced $\chi^2 = 3.19$) was found,
with a best-fit temperature of 6.0 keV. A similarly poor fit was found when
the {\it ROSAT} and {\it ASCA} data were analyzed separately. Poor fits
were also obtained for
blackbody, disk-blackbody, powerlaw, powerlaw with high energy cutoff (CPL),
MEKAL (MKL), a self-Comptonization spectrum after Lamb \& Sanford (1979; CLS),
and a self-Comptonization spectrum after Sunyaev \& Titarchuk (1980) models,
with a reduced $\chi^2$ always exceeding two for $\sim$1065 degrees of freedom.
These models were chosen since they have often been found in the literature to
describe the emission from LMXBs.

To illustrate that
a soft component is needed in addition to the hard component, we have
excluded the {\it ROSAT} data and fit the {\it ASCA} data only in the 2.0--10
keV range with a single component bremsstrahlung model with the absorption
fixed at the Galactic value. Now the fit is good (reduced $\chi^2 = 1.09$)
with $kT=7.4 \pm 0.3$ keV, consistent with the best-fit
bremsstrahlung model obtained by {\it Ginga} over a similar energy range.
We have extrapolated this model down to an energy of 0.2 keV, and plotted
the {\it ROSAT} PSPC data and {\it ASCA} data below 2.0 keV over it
in Figure~\ref{fig:brems}. Although the
7.4 keV bremsstrahlung model provides a reasonable fit to the {\it ROSAT} data
between 1.5--2.4 keV, below 1.5 keV there is a considerable excess of soft
X-ray emission. This same feature is present in the {\it ASCA} data in
the 0.8--1.5 keV range.
A similar exercise using a cut-off powerlaw (CPL) model yielded similar
results, with best-fit values of $1.5 \pm 0.1$ and $12.4^{+6.9}_{-3.3}$ keV
for the powerlaw exponent and cut-off energy, respectively.

\begin{figure*}[htb]
\vskip4.1truein
\includegraphics{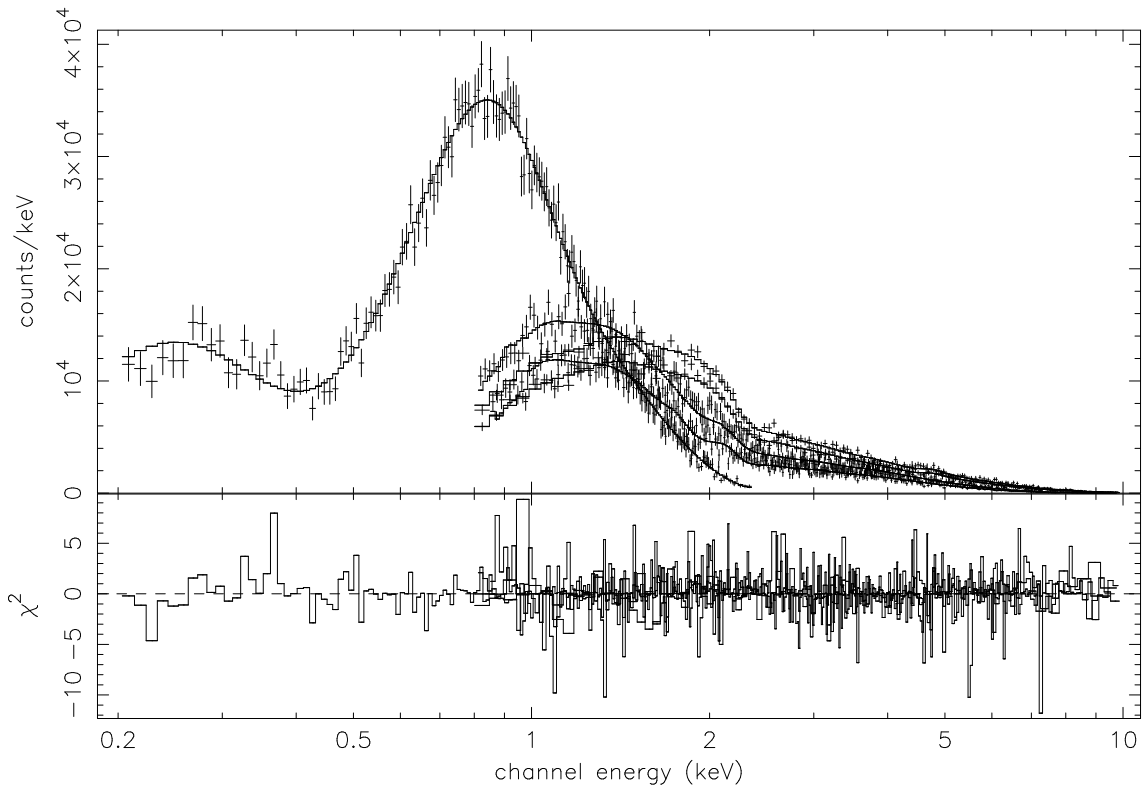}
\caption[MKL+TB]{
{\it ROSAT} PSPC + {\it ASCA} data with best-fit MEKAL+TB model using
data in the 0.2--10 keV bandpass.
\label{fig:mkl+tb}}
\end{figure*}

\vskip0.4truein

\begin{table*}[htb]
\begin{center}
{\sc Table~2}
\vskip0.1truein
\begin{tabular}{lcccccccccc}
\multicolumn{9}{c}{\sc Spectral Fits of M31 Bulge} \cr
\tableline \tableline
&&\multicolumn{2}{c}{Soft Component} &&
\multicolumn{4}{c}{Hard Component} \cr
\cline{3-4} \cline{6-9}
&$N_H$& $kT$&&&$kT$& $\tau$ &&High Energy& \\
~~~~Model& ($10^{20}$ cm$^{-2}$) & (keV) & $Z/Z_{\odot}$
&& (keV) & ($10^{-3}$) & $\Gamma$ & Cutoff (keV) &
$\chi^2_{\nu}$ \\
\tableline
\cr
MKL + TB  & 4.38$^{+0.44}_{-0.27}$ & 0.38$^{+0.03}_{-0.03}$ &
0.20$^{+0.34}_{-0.08}$ && 7.8$^{+0.3}_{-0.3}$ & \ldots
& \ldots & \ldots & 1.14 \\
MKL + TB & 6.73 (fixed) & 0.36$^{+0.03}_{-0.02}$ &
0.04$^{+0.01}_{-0.01}$ && 8.0$^{+0.3}_{-0.3}$ & \ldots
& \ldots & \ldots & 1.23 \\
MKL + CLS & 4.43$^{+0.42}_{-0.48}$ & 0.38$^{+0.04}_{-0.03}$ &
0.18$^{+0.27}_{-0.07}$ && 9.4$^{+1.8}_{-1.7}$ & 1.23$^{+0.21}_{-0.14}$
& \ldots & \ldots & 1.14 \\
MKL + CLS & 6.73 (fixed) & 0.36$^{+0.02}_{-0.02}$ &
0.04$^{+0.01}_{-0.01}$ && 10.3$^{+2.1}_{-1.9}$ & 1.15$^{+0.17}_{-0.15}$
& \ldots & \ldots & 1.22 \\
MKL + CPL & 4.41$^{+0.45}_{-0.48}$ & 0.38$^{+0.04}_{-0.03}$ &
0.21$^{+0.64}_{-0.10}$ && \ldots & \ldots & 1.26$^{+0.07}_{-0.08}$ &
6.86$^{+1.13}_{-0.89}$ & 1.14 \\
MKL + CPL & 6.73 (fixed) & 0.36$^{+0.03}_{-0.02}$ &
0.03$^{+0.01}_{-0.01}$ && \ldots & \ldots & 1.19$^{+0.07}_{-0.09}$ &
6.13$^{+0.93}_{-0.80}$ & 1.22 \\
\tableline
\end{tabular}
\end{center}
\label{tab:spec_fits}
\end{table*}

\subsection{Two Component Models} \label{ssec:two_comp}

Next, various combinations of the spectral models described above were fit
to the data. Whereas the parameters
for the models were linked between the five data sets, the normalizations
of each model were allowed to be independent of one another. The absorption
was left as a free parameter. Of all the possible combinations, only three
gave a fit with a reduced $\chi^2$ less than 1.3 -- the MKL + TB, MKL + CLS,
and MKL + CPL models. The best-fit parameters for
these models are shown in Table~\ref{tab:spec_fits}. The errors given are
90\% confidence levels for one interesting parameter. For all fits there
were approximately 1060 degrees of freedom. The three models gave identical
reduced $\chi^2$ values. Despite the low reduced $\chi^2$ values, the
null hypothesis probability for the fits was quite small (0.1\%) because
of the large number of degrees of freedom in the fit. However, the residuals
had an approximately Gaussian distribution about zero, so although the models
are not formally acceptable, they should lead to accurate fluxes for the
two components. The best-fit MKL + TB model is shown in
Figure~\ref{fig:mkl+tb} along with the residuals. In all three models,
the best-fit absorption was $\sim35\%$
below the Galactic H~{\sc i} value of Stark et al.\ (1992); therefore we have
also fixed the absorption at the Galactic value for these two models, and
determined the best-fit parameters. This caused an increase in the reduced
$\chi^2$ with a very small null hypothesis probability, but still provided
a better fit than
other models with the absorption left as a free parameter. Other than
the metallicity of the MEKAL component, the best-fit
parameters did not change significantly by fixing the absorption at the
Galactic value.

The temperature of the MEKAL component was well-determined (better than
10\% accuracy) with a value
between 0.36--0.38 keV, and a metallicity around 20\% solar when the
absorption was left free, and around 4\% solar when the absorption
was fixed at the Galactic value. For the hard component, the TB model
gave a temperature $7.8 \pm 0.3$ keV. The temperatures are consistent with
the results of the
analysis of a shorter PSPC observation (Irwin \& Sarazin 1998b), but now
the {\it ASCA} data have tightly constrained the temperature of the hard
component, whereas only a lower limit was found before. For the CLS model,
a slightly higher temperature was found along with a low optical depth.
For the CPL model, values of $\Gamma=1.2-1.3$ and a cutoff energy of
6--7 keV are similar to those found for individual Galactic LMXBs
with {\it Einstein} data by Christian \& Swank (1997).
Table~\ref{tab:fluxes} gives the unabsorbed fluxes for the soft and
hard components for each model over various energy ranges.

\begin{table*}[htb]
\label{tab:fluxes}
\begin{center}
{\sc Table~3}
\vskip0.1truein
\begin{tabular}{lccccccccc}
\multicolumn{9}{c}{\sc Unabsorbed Fluxes\tablenotemark{a}} \cr
\tableline \tableline
&&\multicolumn{3}{c}{Soft Component} &&
\multicolumn{3}{c}{Hard Component} \cr
\cline{3-5} \cline{7-9}
&$N_H$&&&&&&& \\
~~~~Model& ($10^{20}$ cm$^{-2}$) & 0.1--2 keV & 0.2--4 keV & 0.25--10 keV
&& 0.1--2 keV & 0.2--4 keV & 0.25--10 keV \\
\tableline
\cr
MKL + TB  & 4.38$^{+0.44}_{-0.27}$ & 0.70$\pm{0.16}$ & 0.59$\pm{0.11}$ &
0.55$\pm{0.09}$ && 1.16$\pm{0.03}$ & 1.73$\pm{0.03}$ & 2.62$\pm{0.03}$ & \\
MKL + TB & 6.73 (fixed) & 1.51$\pm{0.07}$ & 1.14$\pm{0.05}$ &
0.99$\pm{0.04}$ && 1.16$\pm{0.02}$ & 1.72$\pm{0.03}$ & 2.62$\pm{0.03}$ & \\
MKL + CLS & 4.85$^{+0.17}_{-0.37}$ & 0.73$\pm{0.17}$ & 0.61$\pm{0.11}$ &
0.57$\pm{0.10}$ && 1.15$\pm{0.03}$ & 1.72$\pm{0.02}$ & 2.61$\pm{0.02}$ & \\
MKL + CLS & 6.73 (fixed) & 1.53$\pm{0.08}$ & 1.15$\pm{0.05}$ &
1.01$\pm{0.03}$ && 1.14$\pm{0.03}$ & 1.71$\pm{0.02}$ & 2.60$\pm{0.03}$ & \\
MKL + CPL & 4.41$^{+0.45}_{-0.48}$ & 0.69$\pm{0.19}$ & 0.58$\pm{0.14}$ &
0.54$\pm{0.12}$ && 1.19$\pm{0.06}$ & 1.75$\pm{0.05}$ & 2.63$\pm{0.05}$ & \\
MKL + CPL & 6.73 (fixed) & 1.55$\pm{0.09}$ & 1.17$\pm{0.07}$ &
1.02$\pm{0.06}$ && 1.12$\pm{0.06}$ & 1.69$\pm{0.05}$ & 2.58$\pm{0.05}$ & \\
\tableline
\end{tabular}
\end{center}
\tablenotetext{a}{in units of $10^{-11}$ ergs s$^{-1}$ cm$^{-2}$}
\end{table*}

Our results contrast with those of Fabbiano et al.\ (1987), who found no
evidence for a soft component despite the fact that the {\it Einstein} IPC
was sensitive to photon energies down to 0.2 keV. Although the IPC had
poor energy at low energies, a soft component should have been detected.
The soft excess emission becomes apparent around an energy of 1.5 keV
(Figure~\ref{fig:brems}), so energy resolution effects should not have
hindered its detection. We find no obvious
explanation for this discrepancy. The soft component
is seen in multiple {\it ROSAT} PSPC pointings and in all four instruments
of {\it ASCA}. In addition, we analyzed a shorter {\it ASCA} observation
of M31 from the HEASARC archive (60037030) and found similar evidence
for a soft component.

\section{Discussion} \label{sec:discussion}

\subsection{The Origin of the Soft Component} \label{ssec:soft_comp}

From spectral fitting, the need for a soft component in the bulge of M31
is clearly evident. But what is the source of the emission? Two
possibilities exist. First, the soft emission may emanate from a warm
interstellar medium (ISM). The second alternative is that the source
of the soft emission is the same as that of the hard component, namely
LMXBs. The case for each is presented below.

A high resolution {\it ROSAT} HRI study of the bulge of M31 (Primini,
Forman \& Jones 1993) found that 45 point sources within the inner
5$^{\prime}$ of the bulge accounted for 58\% of the bulge emission with
the remaining emission being diffuse. Of the remaining emission, the authors
estimate that about 14\% is the result of the large-angle scattering
component of the point response function of the HRI from the resolved point
sources, and another 15\%--26\% is attributable to the integrated emission
from point sources below the detection threshold, given their derived
luminosity function for the bulge sources. This leaves between 25\%--30\% of
the total emission unexplained by discrete sources or instrumental effects.
Other sources such K and M dwarf stars, cataclysmic variables, and RS CVn stars
were insufficient to explain the remaining diffuse emission.
Primini et al.\ (1993) were forced to conclude that the diffuse emission
is either a new class of X-ray sources or a hot component of the
interstellar medium.

At first glance the interstellar medium explanation seems quite attractive.
From the spectral fits presented here, the soft component 
represents 35\%--40\% of the total X-ray emission in the 0.1--2.0 keV
range (Table~\ref{tab:fluxes}),
at least for the case where the absorption was allowed to vary. This is
roughly the same percentage of the total emission as the unexplained
diffuse emission found by Primini et al.\ (1993). Furthermore, the soft
emission is best-described by a MEKAL model, which is often used to
describe the X-ray emission from a metal-enriched, optically thin
thermal plasma. The stellar velocity dispersion of the bulge of M31
is 151 km s$^{-1}$ (Whitmore 1980). Using the velocity dispersion--X-ray
temperature relation derived from a sample of 30 elliptical galaxies by
Davis \& White (1996) of $kT \propto \sigma^{1.45}$, gas in a potential
well of this magnitude should have a temperature of about 0.3 keV, in
good agreement with the spectral fits. The best-fit metallicity value
is consistent with metallicity measurements obtained from the hot gas in
early-type galaxies (see, e.g., Matsumoto et al.\ 1997).

However, equally compelling pieces of evidence support the claim that
the source of the soft emission is LMXBs. First of all, a similar spatial study
of the bulge of M31 with {\it Einstein} HRI data by Trinchieri \& Fabbiano
(1991) came to a different conclusion than Primini et al.\ (1993) concerning
the nature of the unresolved emission. Trinchieri \& Fabbiano (1991) found
that 75\% of the bulge emission was resolved into 46 point sources (as opposed
to 58\% found by Primini et al.\ 1993 with {\it ROSAT}), with the
remaining diffuse emission easily attributable to point sources below
the detection threshold. Trinchieri \& Fabbiano (1991) also found a somewhat
steeper luminosity distribution function than Primini et al.\ (1993).
{\it Einstein} covered
a harder energy bandpass than did {\it ROSAT}, so if the fainter sources
had harder spectra than the brighter sources, this could lead to a difference
in the measured slope between the two studies. Whether this is the cause
of the discrepancy between the two studies is unclear. Deep observations with
{\it Chandra} will be necessary to determine exactly what percentage of
the total bulge emission cannot be attributed to point sources.

Second, the determination of the contribution of the soft
component relative to the total emission is dependent on the magnitude of
the absorption component assumed. When the absorption was fixed at the
Galactic value, the amount of (unabsorbed) soft flux increased considerably
when compared to the case where the best-fit column density was less than the
Galactic value (Table~\ref{tab:fluxes}). When the column density was fixed at
the Galactic value, the soft component accounted for 55\%--60\% of the total
emission in the {\it ROSAT} (0.1--2.0 keV) band. This is twice the amount
of unexplained diffuse emission found by Primini et al.\ (1993).
Thus, at least half
of the soft emission must emanate from the discrete sources themselves
rather than from any gas present in the bulge, if the model in
which the absorption is fixed at the Galactic value is used.

But perhaps the strongest evidence supporting a soft component emanating
from the LMXBs rather than an interstellar medium lies in the spectra
of the individual LMXBs that have been resolved in the bulge.
Supper et al.\ (1997) analyzed the spectra of 7 of the 22 bulge point sources
resolved by the {\it ROSAT} PSPC and fit them with simple bremsstrahlung
fits. Since the statistics for any one LMXB were poor, this simple model
provided an adequate fit to the spectra. The best-fit
temperatures were in the range 0.45--1.5 keV, well below the canonical
temperature of 5--10 keV previously assumed for LMXBs. This is consistent
with the value of $kT=0.78 \pm 0.07$ keV derived for the bulge of M31 as a whole
with the same model by Irwin \& Sarazin (1998b), although the fit in that
case was only marginal due to better statistics. Nearly all of the
remaining 15 point sources had X-ray colors (ratio of X-ray counts in
three separate energy bands coving the bandpass of the PSPC) that were similar
to the 7 for which temperatures were derived, indicating that they had similar
spectra. Since the bright sources, faint sources, and the total emission
from the bulge all seem to have similar spectra, this strongly
suggests that the soft component seen in the integrated emission from the bulge 
is emanating from the LMXBs themselves, with little emanating for
a hot interstellar medium component.

One final piece of evidence involves comparing the bulge of M31 to the bulge
of the nearby Sa galaxy NGC~1291. Bregman, Hogg, \& Roberts (1995) fit
the {\it ROSAT} PSPC spectrum of NGC~1291 with a hard + soft component
model similar to the one used to fit the bulge of M31. In fact, the X-ray
colors of the bulge of NGC~1291 are nearly identical to those of the bulge
of M31, despite the fact that the 0.5--2.0 keV X-ray--to--optical luminosity
ratio of NGC~1291 is a factor of 1.7 higher than that of the bulge of M31
(Irwin \& Sarazin 1998b). It seems unlikely that the difference in the
$L_X/L_B$ values can be due to there being a higher percentage of the ISM
component in NGC~1291 than in M31; this would lead to a difference in
the X-ray colors between the two galaxies. Irwin \& Sarazin (1998b) found
that the C32 color (defined as the ratio of counts in the 0.91--2.02 keV
band to the counts in the 0.52--0.90 keV band) for M31 and NGC~1291
was $1.16 \pm 0.05$ and $1.15 \pm 0.14$, respectively, after correcting for
absorption (this color is only modestly dependent on absorption). Taking the
MKL+TB spectral model for M31 presented in this paper, we added an additional
MEKAL component to represent an ISM component that might be present in
the spectrum of NGC~1291. This model had a temperature of 0.3 keV and a
metallicity of 30\% solar. This component was added in an amount such that the
0.5--2.0 keV luminosity increased by a factor of 1.7, to represent the
difference in the $L_X/L_B$ values between M31 and NGC~1291. Doing this
caused the C32 value to decrease by more than 30\%. Yet the C32 value for
NGC~1291 was identical to that of M31. If the difference in $L_X/L_B$ is
due in part to an ISM, the ISM component needs to be exactly matched by
an increase in the LMXB component to keep C32 in NGC~1291 the same as
in M31. A more
likely explanation is that the X-ray emission mechanism is identical in
the two galaxies (solely LMXB emission), with NGC~1291 having a higher
percentage of LMXBs per unit optical luminosity than M31.

\subsection{Implications For Early-type Galaxies} \label{ssec:implications}

At some level, stellar X-ray emission must contribute to the total X-ray
emission in early-type galaxies, although that level is yet to be
determined. In X-ray bright (high $L_X/L_B$) galaxies, there is little
doubt that a hot ($\sim$0.8 keV) interstellar medium is responsible for
most of the X-ray emission in these galaxies. But even in these galaxies,
a measurable hard (5--10 keV) component has been detected with {\it ASCA}
(Matsumoto et al.\ 1997), that seems to scale roughly with optical
luminosity and has been attributed to the integrated emission from LMXBs. The
X-ray faint early-type galaxies remain a puzzle. Low X-ray count rates in
these galaxies make them difficult to study, but the emerging picture is
that their X-ray spectra are much different than their X-ray bright
counterparts.

The first piece of evidence that the spectra of X-ray faint early-type galaxies
differed from those of X-ray bright galaxies came from observations performed
by {\it Einstein}. Kim, Fabbiano, \& Trinchieri (1992) found that
X-ray faint galaxies exhibited significant excess very soft X-ray emission.
Subsequent observations using {\it ROSAT} found the X-ray emission of
several X-ray faint galaxies to be described by a two-component (very soft
+ hard) model, with the hard component attributed to LMXBs and the very soft
component of unknown origin (Fabbiano et al.\ 1994;
Pellegrini 1994; Fabbiano \& Schweizer 1995). Irwin \& Sarazin (1998b)
showed this to be the case in all X-ray faint early-type galaxies. The
temperature of the hard component was unconstrained, however, due to the
limited bandpass of {\it ROSAT}.

Kim et al.\ (1996) performed a joint {\it ROSAT} + {\it ASCA} analysis of the
X-ray faint galaxy NGC~4382. The agreement between their derived spectral
parameters for NGC~4382 and the ones presented here for the bulge of M31
are remarkable. Kim et al.\ (1996) found a good fit (reduced $\chi^2$=1.03
for 190 degrees of freedom) with a Raymond-Smith + TB model with variable
absorption. Although we used a MEKAL model instead of a Raymond-Smith
model, the difference between the two models was found to be minimal, affecting
the metallicity the most.
For the Raymond-Smith component Kim et al.\ (1996) found a temperature
of $kT = 0.27-0.41$ keV (90\% confidence) and a metallicity unconstrained
but greater than
10\% at the 90\% confidence level. They found the temperature of the TB
component to be $kT = 4.3-12.8$ keV. A best-fit column density that was
$\sim$$2 \times 10^{20}$ cm$^{-2}$ below the Galactic value was also found
for NGC~4382 as was the case for the bulge of M31. In addition,
the contribution of each
component to the total emission are in good agreement for both galaxies.
Kim et al.\ (1996) found an unabsorbed hard--to--soft flux ratio of
1.5 (1.1--1.9), 2.5 (1.7--2.3), and 3.6 (2.5--4.7) in the 0.1--2 keV,
0.2--4 keV, and 0.25--10 keV bands, respectively (the error ranges were
calculated using the 1$\sigma$ confidence levels on the fluxes given by
Kim et al.\ 1996). From Table~\ref{tab:fluxes}, our flux ratios in those
energy bands are 1.7 (1.3--2.1), 2.9 (2.4--3.4), and 4.8 (4.0--5.6),
respectively (90\% confidence levels).
This agreement in the spectral properties of NGC~4382 and
M31 suggests a common emission mechanism.

The fact that the X-ray spectral properties of M31 and NGC~4382 are
virtually identical, coupled with the fact that no more than 25\%
of the X-ray emission from the bulge of M31 can result
from a warm ISM component points to the interesting (yet not entirely
unexpected) conclusion that LMXBs constitute the majority of the X-ray
emission in X-ray faint early-type galaxies. In these galaxies, it is quite
possible that the ISM has been removed from the galaxy either by
Type Ia supernovae-driven winds, or by environmental effects such as ram
pressure stripping from the intracluster medium through which the
galaxy is moving.

The question of whether the soft component seen in the bulge of M31
is stellar or gaseous will soon be unambiguously answered by {\it Chandra}.
With its excellent spatial resolution, {\it Chandra} will easily
determine if the soft emission is resolved or diffuse. In addition,
{\it Chandra} will be able to resolve point sources in nearby early-type
galaxies, at least those at a distance of Virgo or closer. The results
presented above predict that in both cases the soft component will be
found to emanate primarily from LMXBs.

\acknowledgments

We thank the anonymous referee for useful comments and suggestions.
This research has made use of data obtained through the High Energy
Astrophysics Science Archive Research Center Online Service,
provided by the NASA/Goddard Space Flight Center.
This work has been supported by NASA grant NAG5-3247.

\end{document}